\documentclass[12pt,a4paper]{article}
\usepackage[dvips]{graphicx,color}
\newcommand{\dis}{\displaystyle}

\begin{document}
{\pagestyle{empty}
\vskip 1.0cm

{\renewcommand{\thefootnote}{\fnsymbol{footnote}}
\centerline{\large \bf Generalized-ensemble simulations of spin systems and protein systems}
}
\vskip 1.0cm

\centerline{Takehiro Nagasima$^{a,}$\footnote{\ \ e-mail: nagasima@ims.ac.jp}
Yuji Sugita,$^{a,b,}$\footnote{\ \ e-mail: sugita@ims.ac.jp} 
Ayori Mitsutake,$^{c,}$\footnote{\ \ e-mail: ayori@rk.phys.keio.ac.jp}
and Yuko Okamoto$^{a,b,}$\footnote{\ \ e-mail: okamotoy@ims.ac.jp}}

\vskip 0.5cm

\centerline{$^a${\it Department of Theoretical Studies}}
\centerline{{\it Institute for Molecular Science}}
\centerline{{\it Okazaki, Aichi 444-8585, Japan}}
\centerline{$^b${\it Department of Functional Molecular Science}}
\centerline{{\it The Graduate University for Advanced Studies}}
\centerline{{\it Okazaki, Aichi 444-8585, Japan}}
\centerline{$^c${\it Department of Physics}}
\centerline{{\it Faculty of Science and Technology}}
\centerline{{\it Keio University}}
\centerline{{\it Yokohama, Kanagawa 223-8522, Japan}}

\vskip 1.0cm

\centerline{{\it Computer Physics Communications}, in press.}

\vskip 0.5cm

\centerline{{\it {\bf Keywords:} Potts model; protein folding; 
generalized-ensemble algorithm;}}
\centerline{{\it multicanonical algorithm; simulated tempering; replica-exchange method}}

\medbreak
\vskip 1.0cm
 
\centerline{\bf ABSTRACT}
\vskip 0.3cm

In complex systems such as spin systems and protein systems, 
conventional
simulations in the canonical ensemble will get trapped in states
of energy local minima.  We employ the generalized-ensemble 
algorithms in order to overcome this
multiple-minima problem.  Three
well-known generalized-ensemble algorithms, namely, multicanonical
algorithm, simulated tempering, and replica-exchange method, are
described.  We then present three new generalized-ensemble
algorithms based on the combinations of the three methods.
Effectiveness of the new methods are tested with a Potts model
and protein systems.

 
}


\section{INTRODUCTION}
\label{ssIntro}
The protein folding problem is one of the most challenging
problems in computational biophysics.
The difficulty comes from the fact that
the number of possible conformations 
for each protein is astronomically large.
Simulations by conventional methods such as Monte Carlo (MC) 
or molecular dynamics (MD)
algorithms in canonical ensemble will necessarily get trapped 
in one of many
local-minimum states in the energy function.
In order to overcome this multiple-minima
problem, many methods have been proposed 
(for a review, see, e.g., Ref.~\cite{RevHO2}). 
 
One way to alleviate the difficulty is to perform a simulation
in a {\it generalized ensemble} where
each state is weighted by a non-Boltzmann probability
weight factor so that
a random walk in potential energy space may be realized.
The random walk allows the simulation to escape from any
energy barrier and to sample much wider configurational space than
by conventional methods.
Monitoring the energy in a single simulation run, one can
obtain not only
the global-minimum-energy state but also canonical ensemble
averages as functions of temperature by the single-histogram \cite{FS1}
and multiple-histogram \cite{FSWHAM} reweighting techniques.

One of the most well-known generalized-ensemble methods is perhaps
{\it multicanonical algorithm} (MUCA) \cite{MUCA}
(for a recent review, see Ref.~\cite{MUCArev}).
MUCA was first introduced to the molecular 
simulation field in Ref.~\cite{HO}.
Since then MUCA has been extensively
used in many applications in
protein and related systems (for a review, see, e.g.,
Ref.~\cite{RevHO1}).
  
While a simulation in multicanonical ensemble performs a free
1D random walk in potential energy space, that in
{\it simulated tempering} (ST) \cite{ST} 
performs a free random walk in temperature space
(for a review, see, e.g., Ref.~\cite{STrev}).
This random walk, in turn,
induces a random walk in potential energy space and
allows the simulation to escape from
states of energy local minima.
ST has also been introduced to the protein folding
problem \cite{IRB1,HO97}.

The generalized-ensemble method
is powerful, but in the above two methods the probability
weight factors are not {\it a priori} known and have to be
determined by iterations of short trial simulations.
This process can be non-trivial and very tedius for
complex systems with many local-minimum-energy states.
Therefore, there have been attempts to accelerate
the convergence of the iterative process for MUCA
\cite{SmBr,MUCAW} (see also Ref.~\cite{Landau}).
  
In the {\it replica-exchange method} (REM)
\cite{RE1}, the difficulty of weight factor
determination is greatly alleviated.  (REM is also 
referred to as
{\it multiple Markov chain method} \cite{Whit}
and {\it parallel tempering} \cite{STrev}. 
For recent reviews with detailed references about the method, see, 
e.g., Refs.~\cite{RevMSO,IBArev}.)
In this method, a number of
non-interacting copies (or replicas) of the original system
at different temperatures are
simulated independently and
simultaneously by the conventional MC or MD method. Every few steps,
pairs of replicas are exchanged with a specified transition
probability.
REM has also been introduced to the protein folding 
problem \cite{H97,SO}. 
We further developed a multidimensional REM which is particularly
useful in free energy calculations \cite{SKO}.

However, REM also has a computational difficulty:
As the number of degrees of freedom of the system increases,
the required number of replicas also greatly increases, whereas 
only a single replica is simulated in MUCA or ST.
This demands a lot of computer power for complex systems.
Our solution to this problem is: Use REM for the weight
factor determinations of MUCA or ST, which is much
simpler than previous iterative methods of weight
determinations, and then perform a long MUCA or ST
production run.
The methods are referred to as
the {\it replica-exchange multicanonical algorithm} (REMUCA)
\cite{SO3} and
the
{\it replica-exchange simulated tempering} (REST) \cite{MO4}.
We have introduced a further extension of REMUCA,
 which we refer to as {\it multicanonical replica-exchange method}
 (MUCAREM) \cite{SO3}.
In MUCAREM, the multicanonical weight factor is first
determined as in REMUCA, and then
 a replica-exchange multicanonical production simulation is performed
 with a small number of replicas (for a review of all
 these new methods, see
Ref.~\cite{RevMSO}).

In this article, we describe 
the six generalized-ensemble algorithms
mentioned above.  Namely, we first describe the three familiar methods:
MUCA, ST, and REM.  We then present the three new algorithms:
REMUCA, REST, and MUCAREM.
The effectiveness of these methods is tested with 
a 2-dimensional Potts model and protein systems.

\section{METHODS}
\label{ssMeth}
In the regular canonical ensemble with a given
inverse temperature
$\beta \equiv 1/k_BT$ ($k_B$ is the Boltzmann
constant),
the probability distribution of potential
energy $E$ is given by
\begin{equation}
P_B(E;T)\ ~\propto ~n(E)~ W_B (E;T)~
\equiv \ n(E)~ e^{- \beta E}~,
\label{pb}
\end{equation}
where $n(E)$ is the density of states.
Since the density of states $n(E)$ is a rapidly increasing function of $E$
and the Boltzmann factor $W_B (E;T)$ decreases exponentially with $E$,
the probability distribution $P_B(E;T)$ has a bell-like shape in general.
However, it is very difficult to obtain
canonical
distributions at low temperatures with conventional
simulation methods.
This is because the thermal fluctuations at
low temperatures are small and the
simulation will certainly
get trapped in states of energy local minima.

{\it Multicanonical algorithm} (MUCA) \cite{MUCA}
is one of
the most well-known generalized-ensemble algorithms.
In the ``multicanonical ensemble'' the probability distribution of potential
energy is {\it defined} as follows:
\begin{equation}
P_{mu} (E) ~\propto ~ n (E)~ W_{mu} (E) \equiv {\rm constant}~.
\label{pmu}
\end{equation}
Because the multicanonical weight factor
$W_{mu} (E)$
is (proportional to the inverse of the density of states and)
not {\it a priori} known, one has to
determine it for each system by iterations of trial
simulations.
See, for instance, Ref.~\cite{MUCArev} for details
of the method to
determine the MUCA weight factor
$W_{mu} (E)$.

After the optimal MUCA weight factor is obtained, one
performs a long MUCA simulation once.  By monitoring
the potential energy throughout the simulation, one can find the
global-minimum-energy state.
Moreover, by using the obtained histogram $N_{\rm mu}(E)$ of the
potential energy distribution $P_{\rm mu}(E)$, 
the expectation value of a physical quantity $A$
at any temperature $T=1/k_{\rm B} \beta$
can be calculated from
\begin{equation}
<A>_{T} \ = \frac{\dis{\sum_{E}~A(E)~n(E)~e^{-\beta E}}}
{\dis{\sum_{E} ~n(E)~e^{-\beta E}}}~,
\label{eqn18}
\end{equation}
where the best estimate of the
density of states is given by the single-histogram
reweighting techniques (see Eq.~(\ref{pmu})) \cite{FS1}:
\begin{equation}
 n(E) = \frac{N_{\rm mu}(E)}{W_{\rm mu}(E)}~.
\label{eqn17}
\end{equation}

In {\it simulated tempering} (ST) \cite{ST}
temperature itself becomes a
dynamical variable, and both the configuration and the temperature are updated
during the simulation with a weight:
\begin{equation}
W_{\rm ST}(E;T_m) = e^{-\beta_m E + a_m}~,
\label{Eqn3}
\end{equation}
where we discretize the temperature in
$M$ different values, $T_m$ ($m=1, \cdots, M$).
Without loss of
generality we can order the temperature
so that $T_1 < T_2 < \cdots < T_M$.  The lowest temperature
$T_1$ should be sufficiently low so that the simulation can explore the
global-minimum-energy region, and
the highest temperature $T_M$ should be sufficiently high so that
no trapping in a local-minimum-energy state occurs.
The parameters $a_m$ are chosen so that the probability distribution
of temperature is flat:
\begin{equation}
P_{\rm ST}(T_m) = \int dE~ n(E)~ W_{{\rm ST}} (E;T_m) =
\int dE~ n(E)~ e^{-\beta_m E + a_m} = {\rm constant}~.
\label{Eqn2}
\end{equation}
Hence, in simulated tempering the {\it temperature} is sampled
uniformly. A free random walk in temperature space
is realized, which in turn
induces a random walk in potential energy space and
allows the simulation to escape from
states of energy local minima.

The parameters $a_m$ are not known {\it a priori} and have to be determined
by iterations of short simulations.
See, for instance, Ref.~\cite{HO97} for details
of the method to
determine the ST weight factor
$W_{{\rm ST}} (E;T_m)$.

Note that from Eq.~(\ref{Eqn2}) we have
\begin{equation}
e^{-a_m} \propto \int dE~ n(E)~ e^{- \beta_m E}~.
\label{Eqn4}
\end{equation}
The parameters $a_m$ are therefore ``dimensionless'' Helmholtz free energy
at temperature $T_m$
(i.e., the inverse temperature $\beta_m$ multiplied by
the Helmholtz free energy).

A simulation of ST is realized by alternately
performing the following two steps \cite{ST}.
Step 1:
 A canonical MC or MD simulation at the fixed temperature $T_m$
is carried out for a certain MC or MD steps.
Step 2: The temperature $T_m$ is updated to the 
neighboring values
$T_{m \pm 1}$ with the configuration fixed.  The transition probability of
this temperature-updating
process is given by the Metropolis criterion (see Eq.~(\ref{Eqn3})):
\begin{equation}
w(T_m \rightarrow T_{m \pm 1})
= {\rm min} (1,e^{- \Delta})~,
\label{Eqn5}
\end{equation}
where
\begin{equation}
\Delta = \left(\beta_{m \pm 1} - \beta_m \right) E
- \left(a_{m \pm 1} - a_m \right)~.
\label{Eqn6}
\end{equation}

After the optimal ST weight factor is determined,
one performs a long ST simulation once.
From the results of this production run, one can obtain
the canonical ensemble average of a physical quantity $A$ as a function 
of temperature from Eq.~(\ref{eqn18}), where the 
density of states is given by
the multiple-histogram reweighting techniques \cite{FSWHAM}
as follows.
Let $N_m(E)$ and $n_m$ be respectively
the potential-energy histogram and the total number of
samples obtained at temperature $T_m=1/k_{\rm B} \beta_m$
($m=1, \cdots, M$). 
The best estimate of the density of states is then 
given by \cite{FSWHAM}
\begin{equation}
n(E) = \frac{\dis{\sum_{m=1}^M ~g_m^{-1}~N_m(E)}}
{\dis{\sum_{m=1}^M ~g_m^{-1}~n_m~e^{f_m-\beta_m E}}}~,
\label{Eqn8a}
\end{equation}
where
\begin{equation}
e^{-f_m} = \sum_{E} ~n(E)~e^{-\beta_m E}~.
\label{Eqn8b}
\end{equation}
Here, $g_m = 1 + 2 \tau_m$,
and $\tau_m$ is the integrated
autocorrelation time at temperature $T_m$.
Note that
Eqs.~(\ref{Eqn8a}) and
(\ref{Eqn8b}) are solved self-consistently
by iteration \cite{FSWHAM} to obtain
the dimensionless Helmholtz free energy $f_m$
and the density of states $n(E)$.

The system for {\it replica-exchange method} (REM) \cite{RE1}
consists 
of $M$ non-interacting copies, or replicas, of the original 
system in canonical ensemble at $M$ different 
temperatures $T_m$ ($m=1, \cdots, M$).
  We arrange the replicas so that there is always one replica at each temperature.  
  Then there is a one-to-one correspondence between replicas and temperatures.  
  Let $X = \left\{ \cdots, x_m^{[i]},\cdots \right\}$ stand for a state in this generalized ensemble.  
  Here, the superscript $i$ and the subscript $m$ in $x_m^{[i]}$ label the replica and the temperature, respectively.  
  The state $X$ is specified by the $M$ sets of coordinates $q^{[i]}$.
  A simulation of REM is then realized by alternately performing the following two steps \cite{RE1} (for details of the molecular dynamics
  version of REM, see Ref.~\cite{SO}). 
  Step 1: Each replica in the canonical ensemble at a fixed temperature is simulated simultaneously and independently for a certain number of MC or MD steps.  
  Step 2: A pair of replicas, say $i$ and $j$, which are at neighboring temperatures, say $T_m$ and $T_{m+1}$, 
  respectively,
 are exchanged: $X = \left\{\cdots, x_m^{[i]}, \cdots, x_{m+1}^{[j]}, \cdots \right\} \to X^\prime = \left\{\cdots, x_m^{[j]}, \cdots, x_{m+1}^{[i]}, \cdots \right\}$. 
The transition probability of this replica exchange 
is given by the Metropolis criterion:

\begin{equation}
w(X \rightarrow X^{\prime}) 
= {\rm min} (1,e^{- \Delta})~,
\label{eqn1}
\end{equation}
where
\begin{equation}
\Delta \equiv \left(\beta_{m+1} - \beta_m \right)
              \left(E\left(q^{[i]}\right)
                  - E\left(q^{[j]}\right)\right)~. 
\label{eqn14}
\end{equation}

The {\it replica-exchange multicanonical algorithm} 
(REMUCA) \cite{SO3} and 
{\it replica-exchange simulated tempering} (REST) 
\cite{MO4} overcome
both the difficulties of MUCA and ST (the weight factor
determination is non-trivial)
and REM (a lot of replicas, or computation time, is required).

In REMUCA \cite{SO3} we first perform a short REM 
simulation (with $M$ replicas)
to determine the MUCA
weight factor and then perform with this weight
factor a regular MUCA simulation with high statistics.
The first step is accomplished by the multiple-histogram reweighting
techniques \cite{FSWHAM}.
Let $N_m(E)$ and $n_m$ be respectively
the potential-energy histogram and the total number of
samples obtained at temperature $T_m=1/k_{\rm B} \beta_m$ of the REM run.
The density of states $n(E)$ is then given by solving 
Eqs.~(\ref{Eqn8a}) and (\ref{Eqn8b}) self-consistently by iteration
\cite{FSWHAM}.
Once the estimate of the density of states is obtained, the
multicanonical weight factor can be directly determined from
Eq.~(\ref{pmu}).

In REST \cite{MO4},  
just as in REMUCA,
we first perform a short REM simulation (with $M$ replicas)
to determine the ST
weight factor and then perform with this weight
factor a regular ST simulation with high statistics.
The first step is accomplished by 
the multiple-histogram reweighting
techniques \cite{FSWHAM}, which give
the dimensionless Helmholtz free energy $f_m$ (see Eqs.~(\ref{Eqn8a})
and (\ref{Eqn8b})).
Once the estimate of the dimensionless Helmholtz free energy $f_m$ are
obtained, the simulated tempering 
weight factor can be directly determined by using
Eq.~(\ref{Eqn3}) where we set $a_m = f_m$ (compare Eq.~(\ref{Eqn4})
with Eq.~(\ref{Eqn8b})).

The formulations of REMUCA and REST are simple and straightforward, but
the numerical improvement is great, because the weight factor
determination for MUCA and ST becomes very difficult
by the usual iterative processes for complex systems.

While multicanonical simulations are 
usually based on local
updates, a replica-exchange process can be considered to be a
global update, and global updates enhance the sampling further.
Here, we present a further modification of REMUCA and refer to the
new method as {\it multicanonical replica-exchange method}
(MUCAREM) \cite{SO3}.  In MUCAREM the final production run is not a 
regular multicanonical simulation but a replica-exchange simulation
with a few replicas
in the multicanonical ensemble. 
Because multicanonical simulations cover much wider energy
ranges than regular canonical simulations, the number of
required replicas for the production run of MUCAREM is
much less than that for the regular REM,
and we can keep the merits of
REMUCA (and improve the sampling further).

\begin{figure}[hbtp]
\begin{center}
\includegraphics[width=12.0cm,keepaspectratio]{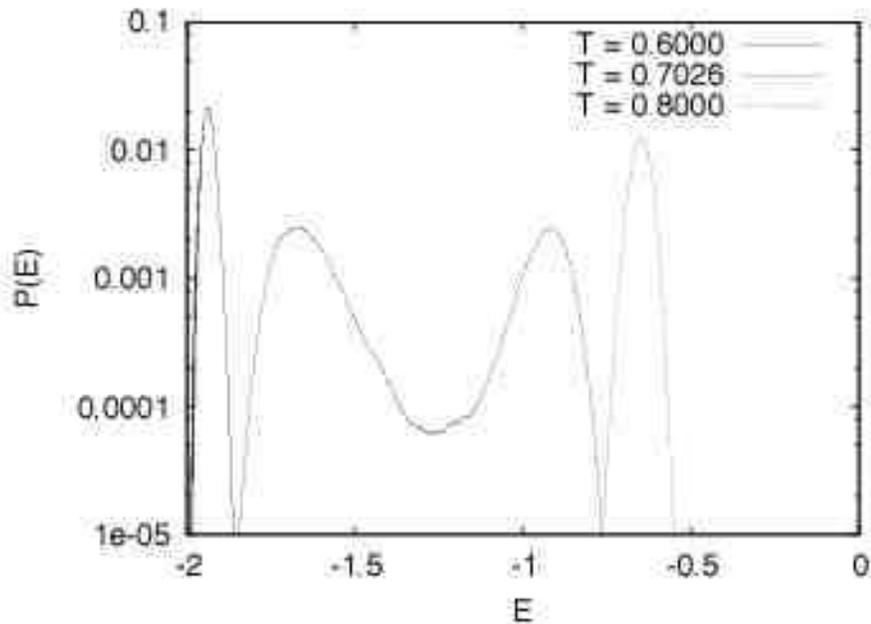}
\end{center}
\caption{Probability distributions of energy of 2-dimensional
10-state Potts model at three temperatures: $T=0.6000$,
0.7026, and 0.8000.  The results were obtained from a
multicanonical MC simulation.}
\label{fig1}
\end{figure}

\section{RESULTS}
\label{ssResDis}
We now present the results of our simulations
based on 
the algorithms described in the previous section.

\begin{figure}[hbtp]
\begin{center}
\includegraphics[width=7.0cm,keepaspectratio]{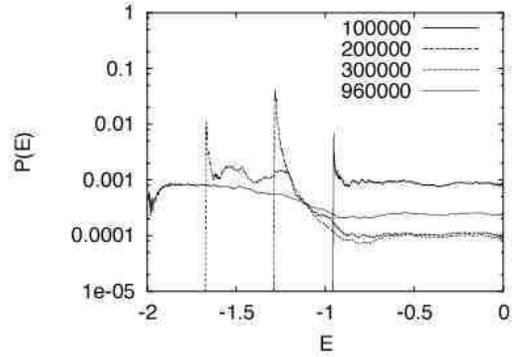}
\end{center}
\caption{Iterative process of multicanonical weight factor determination
for the 2-dimensional
10-state Potts model.
The results after 100,000 MC sweeps, 200,000 MC sweeps, 300,000
MC sweeps, and 960,000 MC sweeps are superimposed.}
\label{fig2}
\end{figure}

\begin{figure}[hbtp]
\begin{center}
\includegraphics[width=7.0cm,keepaspectratio]{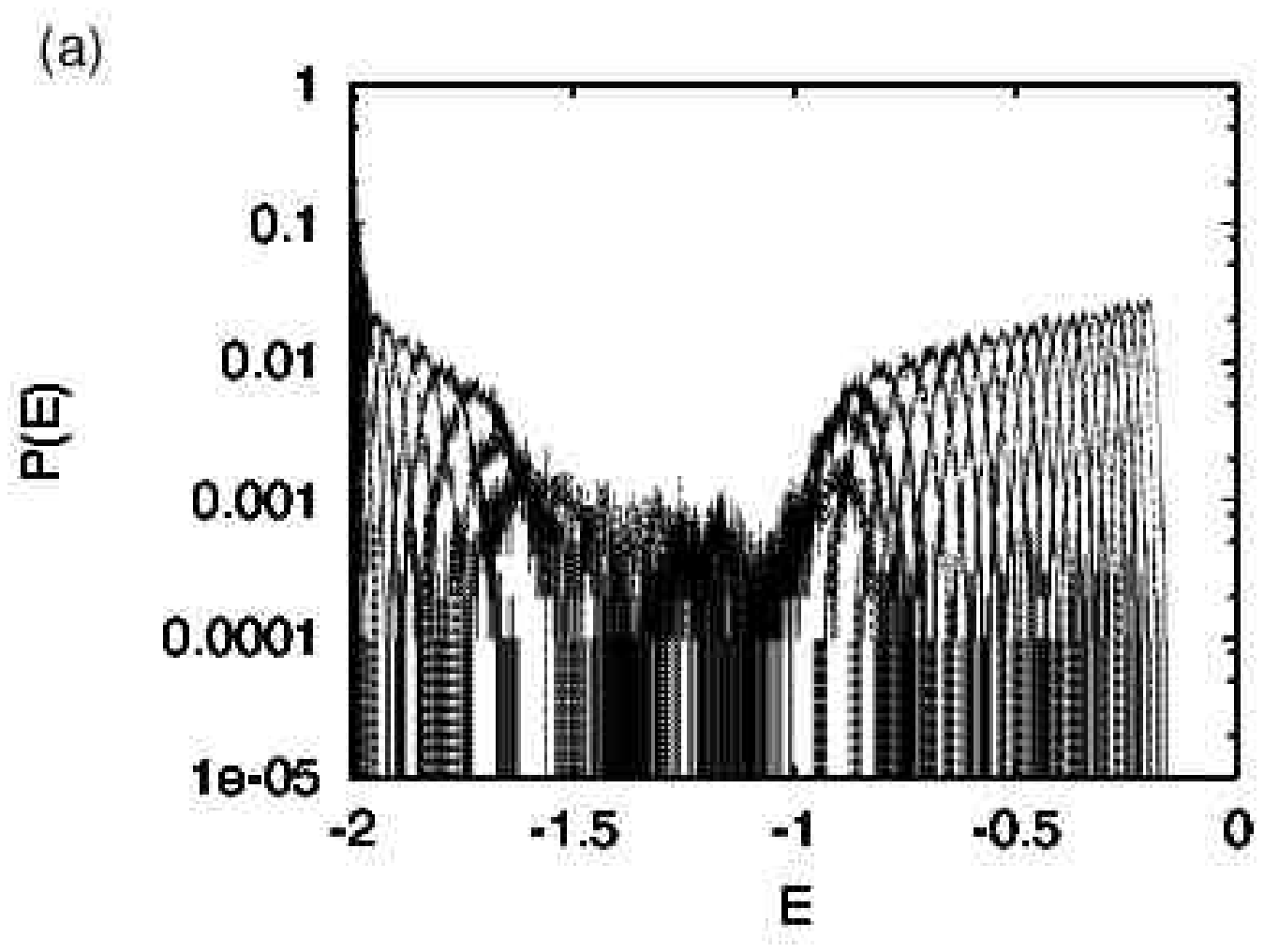}
\includegraphics[width=7.0cm,keepaspectratio]{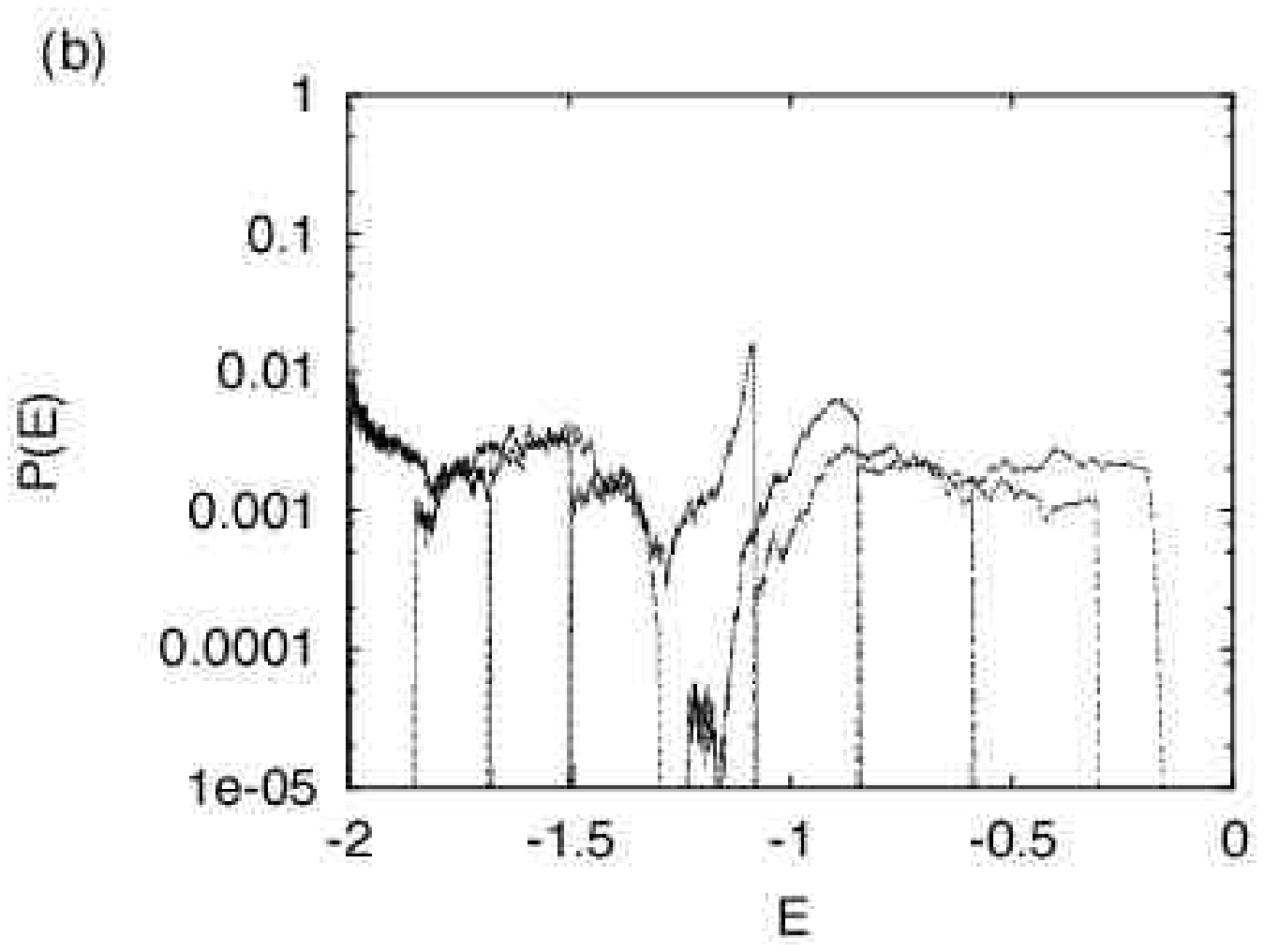}
\includegraphics[width=7.0cm,keepaspectratio]{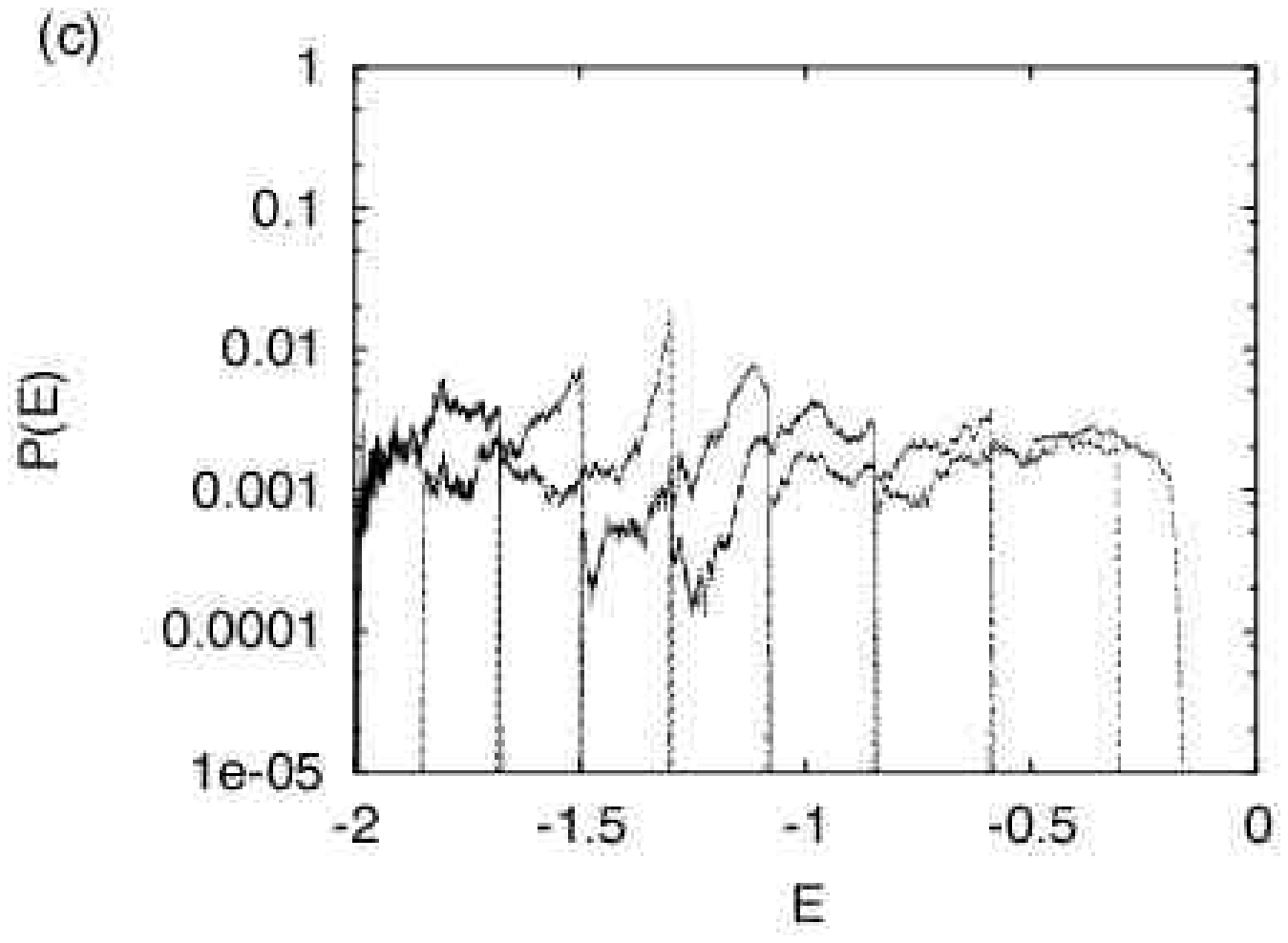}
\includegraphics[width=7.0cm,keepaspectratio]{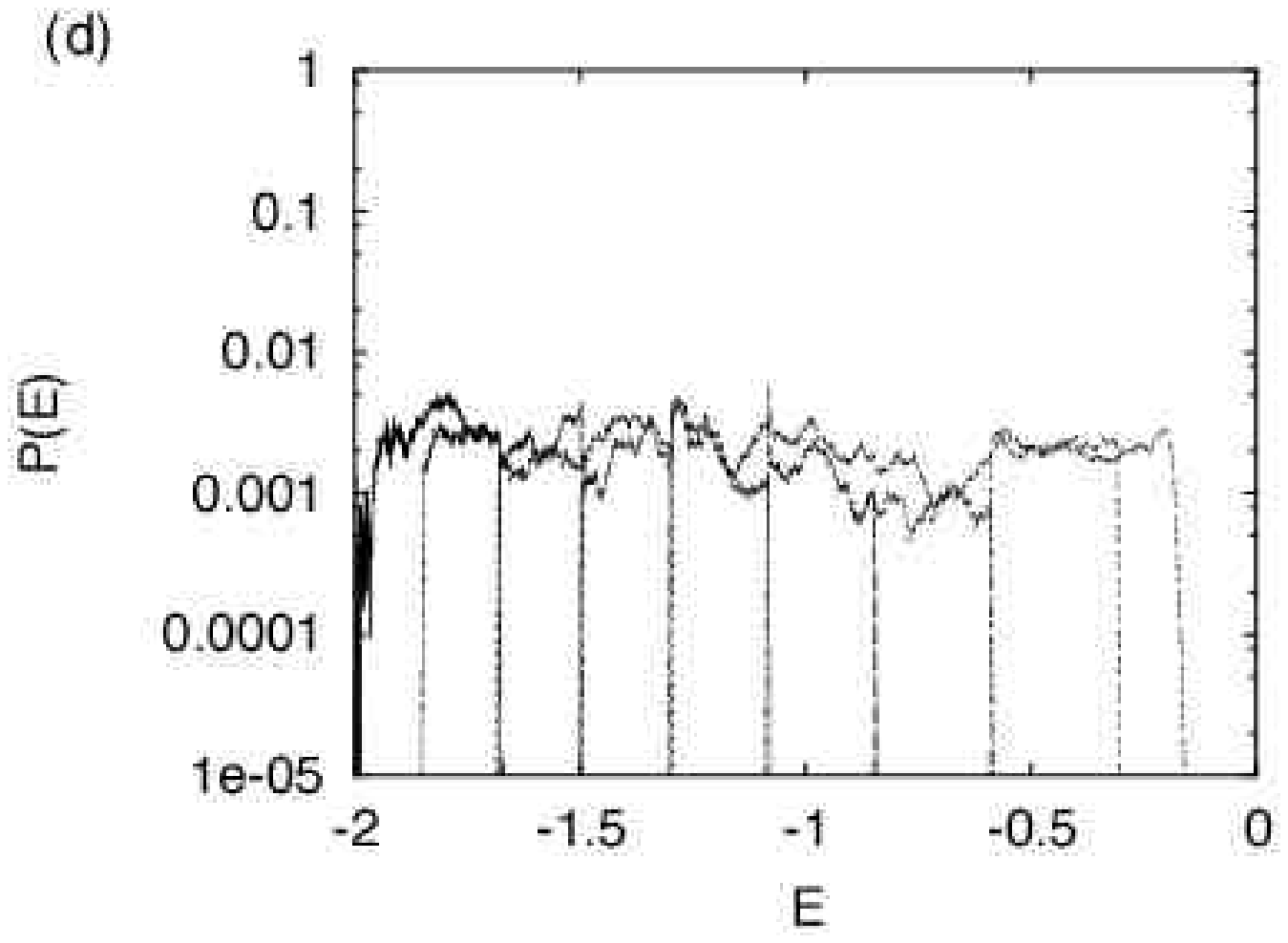}
\end{center}
\caption{Probability distributions of energy
for the 2-dimensional 10-state Potts model:
the results of REM simulation with 32 replicas (a) and
iterations of MUCAREM simulations with 8 replicas (b),
(c), and (d).}
\label{fig3}
\end{figure}

The first example is a spin system.
We studied the 2-dimensional 10-state Potts model
\cite{NSMO}.
The lattice size was $34 \times 34$.
This system exhibits a first-order phase transition
\cite{Bax}.
In Figure 1 we show the probability distributions
of energy at three tempeartures (above the critical
temperature $T_C$, at $T_C$, and
below $T_C$).  At the critical temperature
we observe two peaks in the distribution, indicating
that the system indeed undergoes a first-order phase transition.

In Figure 2 we show how the iterative procedure
\cite{MUCAW}
for the MUCA weight factor determination
converges.  We see that a flat distribution in the entire energy
range was obtained after 960,000 MC sweeps.  Note that
the convergence slows down drastically near the 
global-minimum-energy region (step from 300,000 MC sweeps to
960,000 MC sweeps).

\begin{figure}[hbtp]
\begin{center}
\includegraphics[width=7.0cm,keepaspectratio]{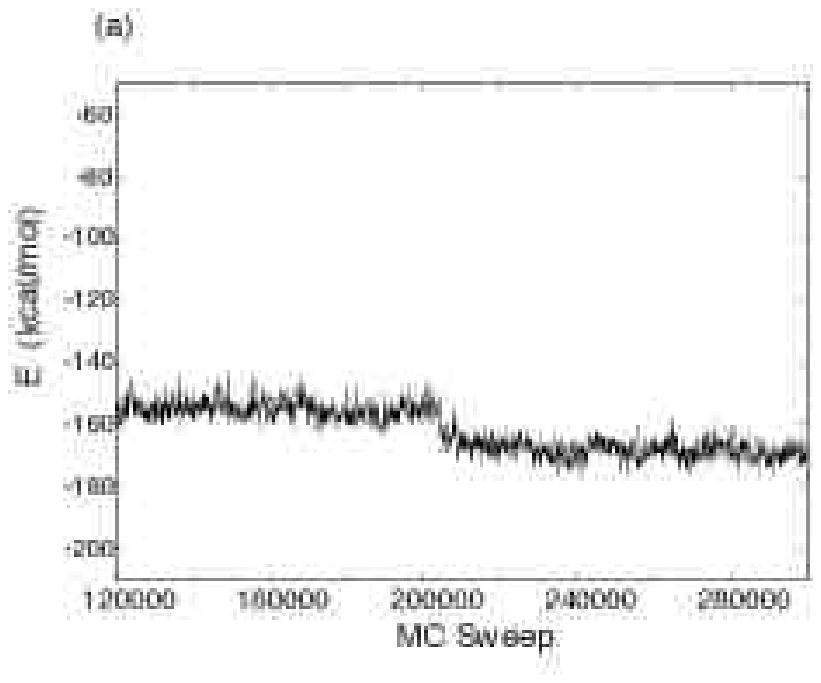}
\includegraphics[width=7.0cm,keepaspectratio]{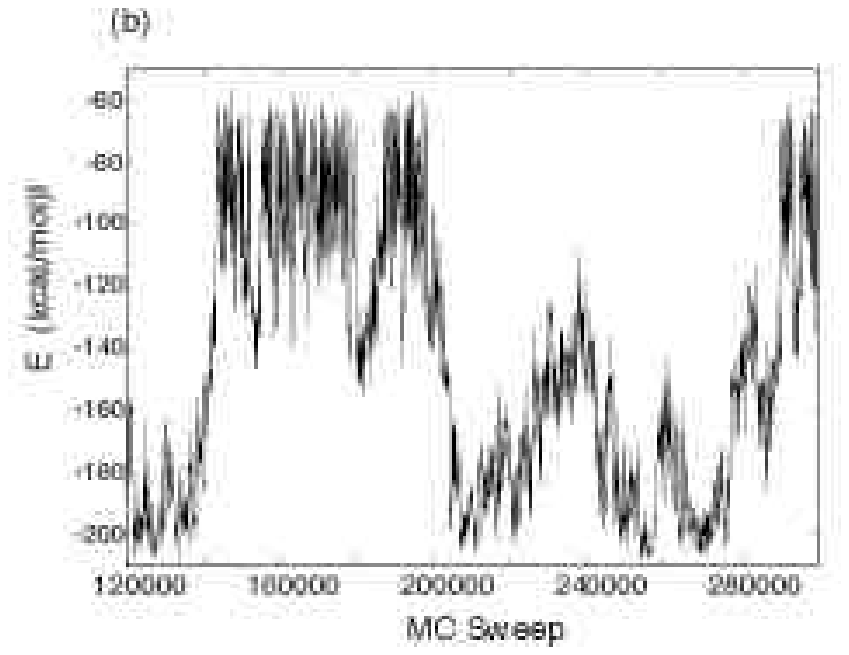}
\end{center}
\caption{Time series (from 120,000 MC sweeps to
300,000 MC sweeps) of potential energy of the
peptide fragment of ribonuclease T1 from
a conventional canonical MC simulation at $T=200$ K
(a) and a multicanonical MC simulation (b).}
\label{fig4}
\end{figure}

In Figure 3 we show the results of our new method for the MUCA
weight factor determination.  We first made a REM
simulation of 10,000 MC sweeps (for each replica)
with 32 replicas (Figure 3(a)).  
Using the obtained energy
distributions, we determined the (preliminary)
MUCA weight factor by the REMUCA procedure as described in
the previous section.
Because the trials of
replica exchange are not accepted near the critical
temperature for first-order phase transitions, the probability 
distributions in Figure 3(a) for the energy range from
$\sim - 1.5$ to $\sim -1.0$ fails to have sufficient overlap,
which is required for successful application of REM.
This means that the MUCA weight factor, or density of states,
in this energy range thus determined is of ``poor quality.''
With this MUCA weight factor, however, we made iterations of
three MUCAREM simulations of 10,000 MC sweeps (for each
replica) with 8 replicas (Figures 3(b), 3(c), 3(d)).
In Figure 3(b) we see that the distributions are not
completely flat, reflecting the poor quality in the
phase-transition region.  This problem is rapidly
rectified as iterations continue, and the distributions
are completely flat in Figure 3(d), which gives an
optimal MUCA weight factor in the entire energy range.
The details including the comparisons with the
new method in Ref.~\cite{Landau}
will be published elsewhere \cite{NSMO}.

\begin{figure}[hbtp]
\begin{center}
\includegraphics[width=7.0cm,keepaspectratio]{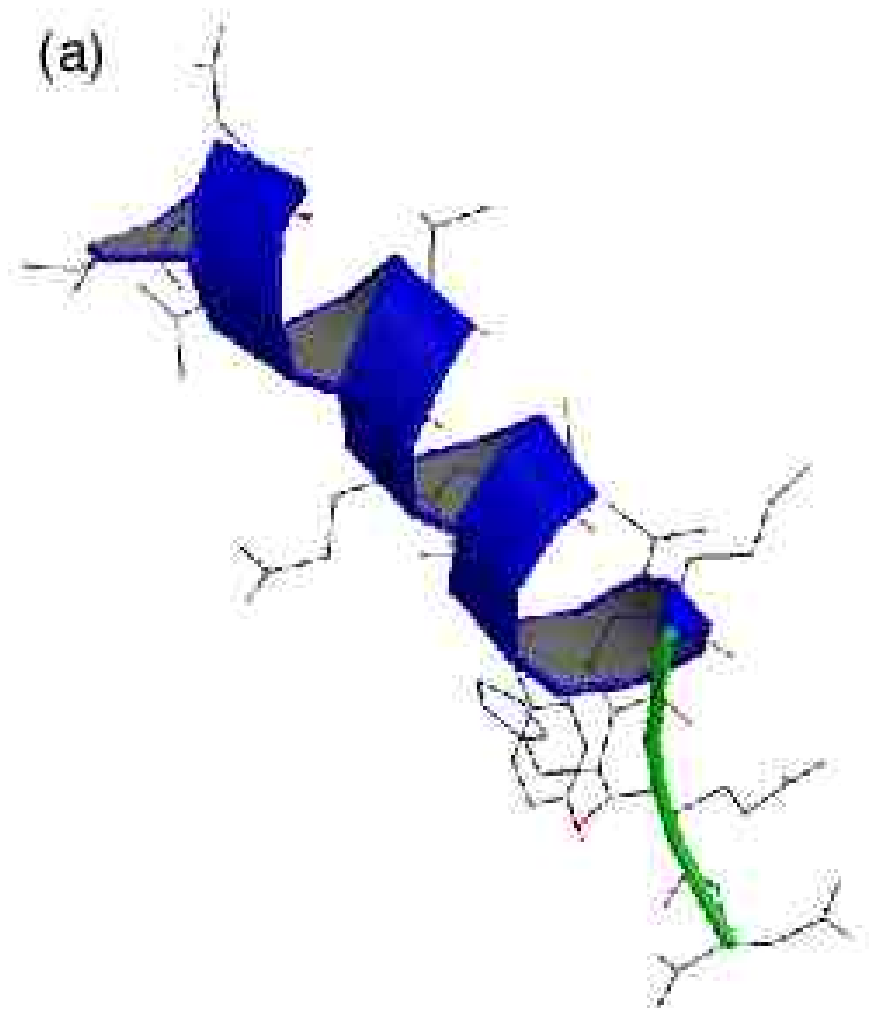}
\includegraphics[width=7.0cm,keepaspectratio]{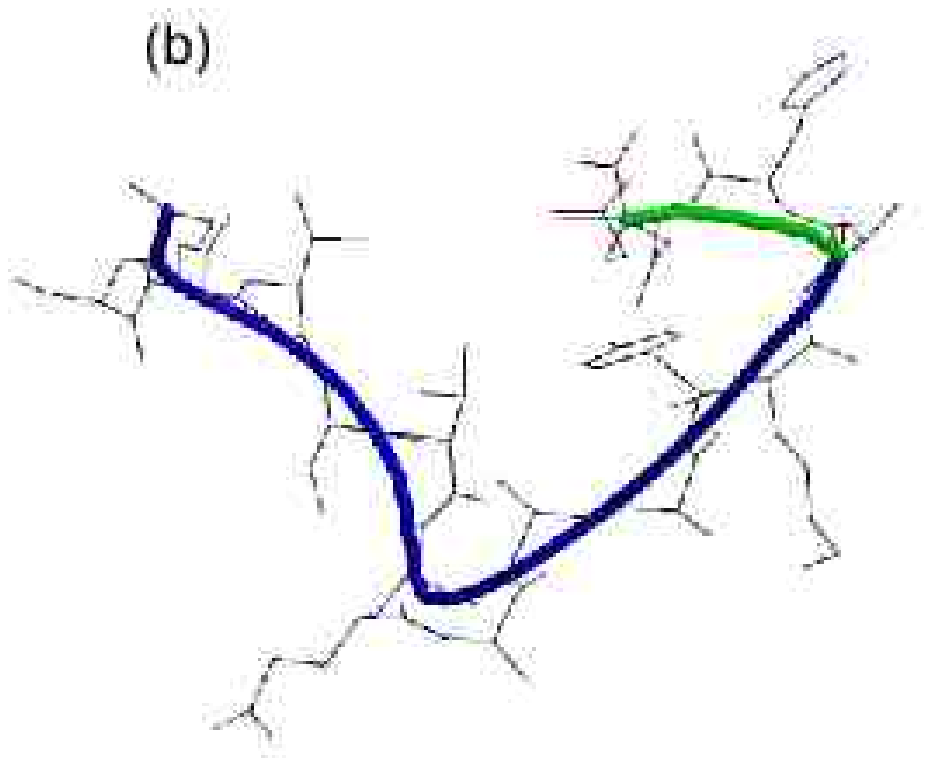}
\includegraphics[width=7.0cm,keepaspectratio]{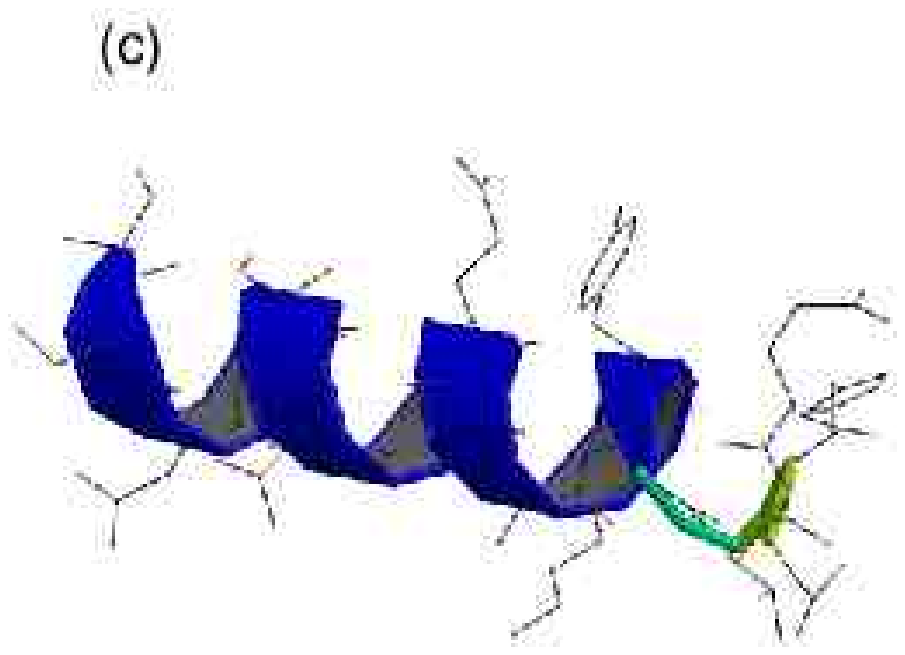}
\includegraphics[width=7.0cm,keepaspectratio]{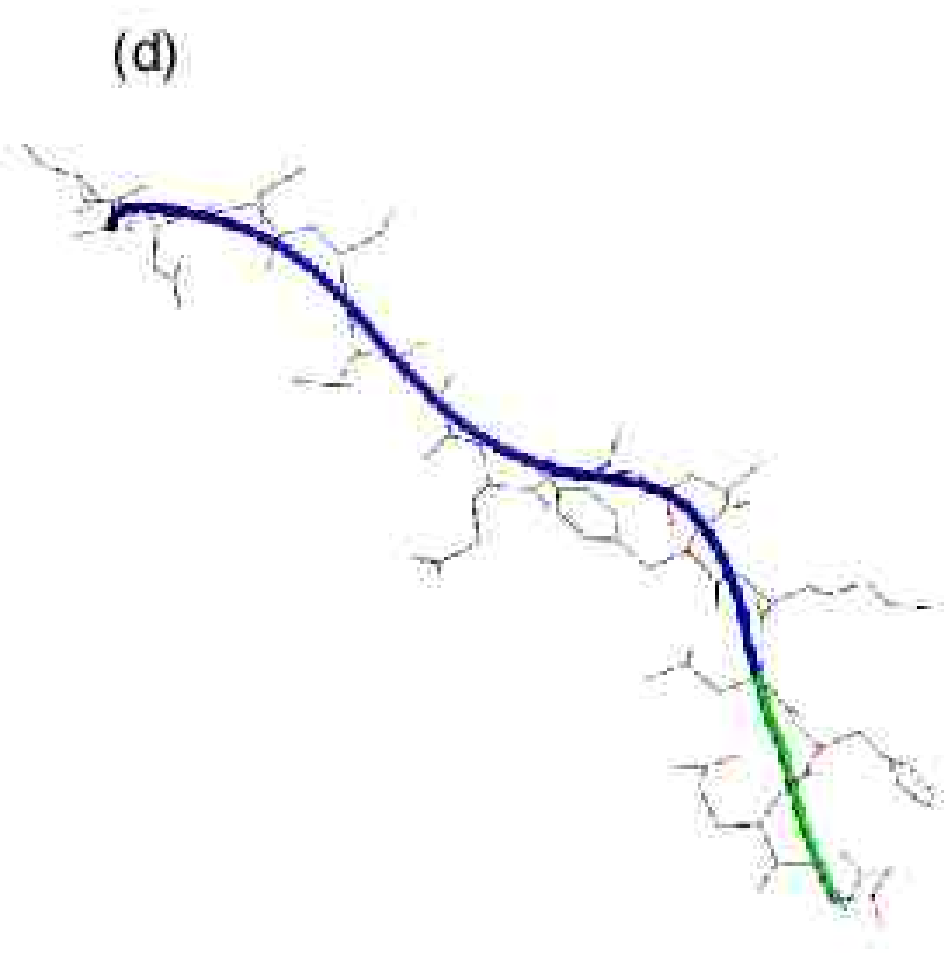}
\end{center}
\caption{Typical snapshots from the multicanonical MC simulation of Figure 4(b).  The corresponding numbers of MC sweeps are 138,000 (a), 190,000 (b),
243,000 (c), and 295,000 (d).  The figures were created with 
Molscript and Raster3D.}
\label{fig5}
\end{figure}

\begin{figure}[hbtp]
\begin{center}
\includegraphics[width=10.0cm,keepaspectratio]{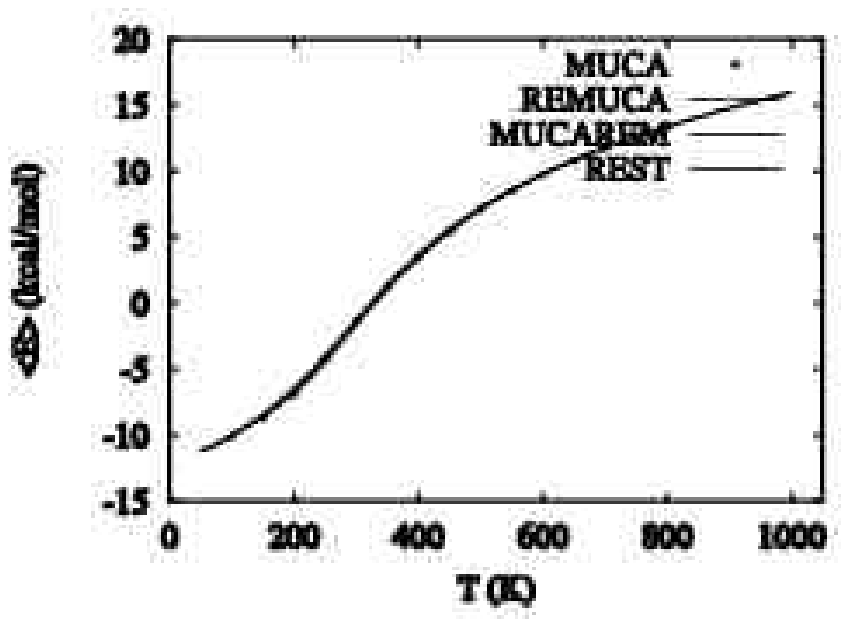}
\end{center}
\caption{The average potential energy of Met-enkephalin
  in gas phase as a function of temperature.
  The results from the four generalized-ensemble algorithms,
  MUCA, REMUCA, MUCAREM, and REST, are superimposed.}
\label{fig6}
\end{figure}

The second example is a protein system.
We first illustrate how effectively generalized-ensemble
simulations can sample the configurational space compared
to the conventional simulations in the canonical ensemble.
It is known by experiments that the system 
of a 17-residue peptide fragment from 
ribonuclease T1 tends to form $\alpha$-helical
conformations.  We have performed both a
canonical MC simulation of this peptide at a low 
temperature ($T=200$ K) and a
multicanonical MC simulation \cite{MO5}.  In Figure 4
we show the time series of potential energy from these
simulations.

We see that the canonical
simulation thermalizes very slowly.
On the other hand, the MUCA simulation indeed performs
a random walk in potential energy space covering a
very wide energy range.  Four conformations
chosen during this period (from 120,000 MC sweeps to
300,000 MC sweeps) are shown 
in Figure 5 for the MUCA simulation.
The MUCA simulation indeed samples a wide 
conformational space.

The last example is a penta peptide, Met-enkephalin,
whose amino-acid sequence is: Tyr-Gly-Gly-Phe-Met.
In Figure 6, we show the 
average
potential energy of Met-enkephalin in gas phase
as a function of temperature that
was calculated by the single- and multiple-histogram
reweighting techniques
from the four generalized-ensemble
algorithms, MUCA, REMUCA, MUCAREM, and REST \cite{MSO}.
The results are in good agreement.

\section{CONCLUSIONS}
\label{ssConc}
In this article we have described the 
formulations of the three well-known
generalized-ensemble algorithms, namely,
multicanonical algorithm (MUCA),
simulated tempering (ST), and replica-exchange method (REM).
We then introduced three new generalized-ensemble
algorithms that combine the merits of the above three methods, 
which we refer to as replica-exchange multicanonical
algorithm (REMUCA), replica-exchange simulated tempering (REST),
and multicanonical replica-exchange method (MUCAREM).

With these new methods available,
we believe that we now have working simulation algorithms
for spin systems and protein systems.

\section{Acknowledgements}
\label{ssAck}
  Our simulations were performed on the HITACHI and other computers at the Research Center for Computational Science, Okazaki National Research Institutes.
  This work is supported, in part, by a grant from the Research for the Future Program of the Japan Society for the Promotion of Science (JSPS-RFTF98P01101).

\end{document}